\documentstyle[preprint,aps]{revtex}
\begin{document}

\def\pmb#1{\setbox0=\hbox{#1}%
  \kern-.025em\copy0\kern-\wd0
  \kern.05em\copy0\kern-\wd0
  \kern-.025em\raise.0433em\box0 }
\def\grad{{\pmb{$\nabla$}}}

\newcommand{\figref}[1]{Fig.\ #1}
\newcommand{\dotprod}{\thinspace \raise2pt \hbox{.}}

\draft

\title{Simulation of the Zero Temperature Behavior of a 3-Dimensional
	 Elastic Medium}

\author{David McNamara}
\author{A. Alan Middleton}
\address{Physics Department, Syracuse University, Syracuse, NY 13244}
\author{Chen Zeng}
\address{Department of Physics and Astronomy, 
Rutgers University, Piscataway, NY 08854}

\date{\today}

\maketitle

\begin{abstract}

We have performed numerical simulation of a 3-dimensional elastic medium, 
with scalar displacements, subject to quenched disorder.  
In the absence of topological
defects this system is equivalent to a 
$(3+1)$-dimensional interface subject to a periodic pinning potential.
We have applied an efficient combinatorial
optimization algorithm to generate exact ground states for 
this interface representation.  Our results
indicate that this Bragg glass is characterized by power law divergences
in the structure factor $S(k)\sim A k^{-3}$.
We have found numerically consistent values of the coefficient $A$ 
for two lattice discretizations of the medium, supporting
universality for $A$ in the isotropic systems considered here.
We also examine the response of the ground state to the change in boundary
conditions that corresponds to introducing a single 
dislocation loop encircling the
system.  The rearrangement of the ground state caused by this change 
is equivalent to the domain wall of elastic deformations which 
span the dislocation loop.
Our results indicate that these domain walls
are highly convoluted, with a fractal dimension $d_f=2.60(5)$.
We also discuss the implications of the domain wall energetics
for the stability of the Bragg glass phase.
Elastic excitations similar to these domain walls
arise when the pinning potential
is slightly perturbed.  As in other disordered systems,
perturbations of relative
strength $\delta$ introduce a new length
scale $L^* \sim \delta^{-1/\zeta}$ beyond which 
the perturbed ground state becomes uncorrelated with the reference
(unperturbed) ground state.
We have performed scaling analysis of the response of the
ground state to the perturbations and obtain $\zeta = 0.385(40)$.
This value is consistent with 
the scaling relation $\zeta=d_f/2- \theta$, where 
$\theta$ characterizes
the scaling of the energy fluctuations of low energy excitations.
\end{abstract}


\widetext
Observation of glassy behavior in flux line arrays in high $T_c$
superconductors \cite{glassySC} calls for a thorough theoretical 
description of such behavior. 
In this system the collective pinning of the flux line array,
rather than the interactions of a single flux line with the
disorder, can dominate the physics \cite{ffh}.
For weak pinning, where dislocations are believed to be
unimportant at large length scales \cite{Fisher,gLeD}, the
entire flux line array can be modeled as a single medium
subject to a pinning potential.
Analytic calculations carried out using the approximation of 
linear elasticity and including the effects of the 
short range order in this system
indicate that quasi-long range order exists in 3 dimensions 
\cite{gLeD,CO,Villain}.  
The elastic medium  assumption was justified 
{\it a posteriori} and is further supported
by an approximate domain wall renormalization calculation \cite{Fisher}. 
The structure factor of a topologically ordered system 
was predicted by these calculations to have power 
law divergences of the form $S(k) \sim k^{-3}$.
We will consider only the case of scalar displacements, which
also models a charge density wave pinned by
charge impurities \cite{CDW}.

We have numerically generated ground states for an elastic medium
subject to quenched point disorder in the 
topologically ordered phase.  Our results for the coefficient
of the divergence of
$S(k)$ lie between the renormalization group and 
Gaussian variational method results obtained by 
Giamarchi and Le Doussal \cite{gLeD}.  
In addition to supporting their analysis, we are able to examine the
response of the system to changes in boundary conditions 
and pinning potential.
By a suitable choice of the boundary conditions we can simulate 
the domain wall of elastic deformations induced by a dislocation 
loop \cite{Fisher}.
The energy of the domain wall dominates the random part of the energy cost
of introducing a single topological defect \cite{gingrashuse}.
Our results
on the energetics of the domain walls thereby indirectly support the 
analysis carried out by Fisher \cite{Fisher}, which indicated that the
this system is marginally stable with respect to the introduction
of dislocations.  
The numerically generated domain walls were found to have a fractal 
dimension $d_f=2.60(5)$.  
At large length scales
the ground state is highly
sensitive to small perturbations in the disorder potential,
as in spin glasses and other disordered systems
{\cite{spin glass,dprm,2delas}}.  
Perturbations of relative strength
$\delta$ in the disorder decorrelate the ground state on length
scales $L^* \sim \delta ^{-1/\zeta}$, with $\zeta = 0.385(40)$.
We are able to relate this response to disorder perturbations to
the properties of the domain walls.

We have generated exact ground states for a discrete model whose
energy in the continuum limit is given by
\begin{equation}
\label{eq:Hamiltonian}
H=\int d^3x \ {c\over 2} \thinspace [\nabla u(\vec x) ]^2 + 
  V(u(\vec x),\vec x)
\end{equation}
with distortions of the medium
represented by $u(\vec x)$, which is assumed to be slowly varying
over the system.  The coefficient $c$ is the elastic constant.
The potential felt by the medium
due to the randomly placed impurities is represented by 
$V(u(\vec x),\vec x)$.
In microscopic descriptions of an elastic system subject to weak disorder
there is a length scale $\xi_p$ below which the elastic energy
dominates and the medium is ordered.  
This short range order manifests
itself here as correlations in the disorder potential 
of the form $V(u,\vec x)=V(u+a , \vec x)$, with $a$ the intrinsic
period of the medium.  The period of the potential, $a$, is the 
lattice spacing in a flux line array or the wavelength of
a charge density wave.
Although this Hamiltonian is insufficient to describe the 
core of a dislocation loop, it can serve to describe the
sheet of elastic deformations which span the loop, 
since the approximation of linear elasticity
breaks down only in the region near the dislocation core.
Comparing the ground state
of Eq.~(\ref{eq:Hamiltonian}) in a system of size $L$ 
subject to periodic boundary conditions
to that for the same disorder realization with twisted 
boundary conditions in one direction (i.e., $u(x,y,0)=u(x,y,L)+a$) 
allows for the identification of the domain wall which would be
caused by a single
dislocation loop encircling the system \cite{Fisher}.

The elastic Hamiltonian, Eq.~(\ref{eq:Hamiltonian}), describes 
a $(3+1)$-dimensional interface subject to a disorder potential.  
In this picture, the displacement variable $u(\vec x)$
maps to the height of the directed interface.  
This interface model has a natural discrete representation in which 
the configuration of the interface is specified by the set of 
bonds it cuts in a 4-dimensional lattice \cite{lattice pinning}.  The
bonds of this lattice are
assigned weights which directly correspond to the
disorder potential.  The sum of the weights of the bonds which the
interface cuts
give the energy of the configuration corresponding to the disorder
energy $\int d^3x  \ V(u,\vec x)$.  
In these discrete models an effective elastic constant arises 
from a dependence of the number of configurations on the average 
gradient of the interface.
Maximal flow algorithms \cite{graphAlgo}, a subclass of 
combinatorial optimization algorithms, allow for the
generation of ground states of this discrete representation
of the interface \cite{middle}.

The lattices numerically studied here are
composed of $L^3\times U$ nodes, where $L$ is
the linear size of the elastic medium, 
and $U$ is the extent of the lattice in the $\hat u$ direction
in which the displacement variable fluctuates.  
Unlike the simulation  of elastic manifolds \cite{middle}, where the
bond weights are non-periodic, there are long range
correlations in the disorder in the $\hat u$ direction.  
We generated random integer weights, chosen from a uniform distribution
over $[0,V_{max}]$, independently for each of the forward bonds
in a layer of unit cells at constant height.
The data we will present were obtained with $V_{max}=5000$, but
we have verified that
our results on the structure of the interface are not significantly altered
for $V_{max}$ as low as $100$.  Throughout the following discussion,
we have normalized the energy of the system, so that the effective 
range of the bond weights is in $[0,1]$.
This set of bond weights on one layer is sufficient to fix
the value of the disorder
on all of the bonds because we require that bonds which differ only by
translations along $\hat u$ have the same weight.
In order to study the universality of the coefficient of the divergence
in $S(\vec k)$, we have simulated interfaces on the
simple hypercubic lattice (SHC) 
as well as the Z-centered hypercubic lattice (ZHC) \cite{4d lattices}. 
For the SHC lattice $\hat u$ was chosen to
be along the (1111) crystallographic axis.  The elementary bonds 
in this lattice are: $(x, y, z, u)=$ 
$\pm (0,\sqrt{2}/2,1/2,1/2)$, $\pm (0,-\sqrt{2}/2,1/2,1/2)$, 
$\pm (\sqrt{2}/2,0,-1/2,1/2)$, $\pm (-\sqrt{2}/2,0,-1/2,1/2)$.
In the ZHC lattice we considered the following
12 bonds extending from
each node: $(x, y, z, u)=$
$(\pm \sqrt{2}/2,0,0,\pm \sqrt{2}/2)$, 
$(0,\pm \sqrt{2}/2,0,\pm \sqrt{2}/2)$,
$(0,0,\pm \sqrt{2}/2,\pm \sqrt{2}/2)$ \cite{ZC}.  
These lattices are the natural extensions of the two types of lattices used
in simulations of the ground state of a 2-dimensional 
elastic medium \cite{dis sub}. Both
types of lattices were simulated using periodic boundary conditions
in the transverse directions.  In addition, in the ZHC lattice
the ground state was computed for each realization of disorder with twisted
boundary conditions in one direction.

We have developed a custom implementation of the push-relabel
maximal flow algorithm \cite{pr}
optimized for application to the regular lattices considered here.
Our modifications to the Push-Relabel algorithm reduced
the memory requirements by nearly a
factor of ten, allowing for simulation of systems composed of up to
approximately
$6\times 10^6$ nodes using less than $512$MB.  
The primary modification involves computing nearest neighbor relations
as needed rather than storing this information.  Overhangs in the interface
are precluded by assigning a large weight to backwards arcs \cite{middle}.
Since these backwards arcs have an effectively infinite weight, 
they cannot be be part of the minimal cut.  
Thus our algorithm can operate without storing
their weight, provided that flow is always allowed to move along the 
backwards arcs.  These modifications
increased the running time of the algorithm, but
obtaining the ground state for each
realization of disorder still took less than one hour
of processor time on a single
400 MHz Pentium II CPU for the largest system sizes studied.  The
memory requirement is linear 
in the number of nodes $N$;
the processor time was found to scale approximately as $N^{1.3}$, 
compared with the worst case bound of $N^2$ \cite{pr}.

The modest computational requirements of this algorithm have allowed us 
to average the properties of the ground states for a variety of system 
sizes over a large number of disorder realizations.  
In addition to generating 
the value of the minimal energy, the algorithm produces the configuration
of the interface.  The interface can be then represented by $u(\vec x)$, 
which is defined on the 3-dimensional lattice formed by projecting the 
interface along the $\hat u$ direction \cite{lattice types}.  
Due to the periodicity of 
the disorder, the energy is invariant under global translations 
of integer multiples of $a$ in the displacement variable
$u(\vec x) \rightarrow u(\vec x)+na$.
Considering the set of the forward bonds cut at 
each location provides a characterization of the
configuration equivalent to measuring the gradient of the interface.
This representation of the interface is useful when comparing
different ground states since it insensitive to
global shifts of $u$.
For the SHC lattice, we have generated  at least $10^3$ realizations
of disorder for systems of size $L=8, 16, 24, 32, 40, 48, 60, 80$.  
We chose the extent of the lattice
in the displacement direction $\hat u$ to ensure that the
boundaries of the system do 
not affect the ground state.  For the SHC lattice $U=20$ was sufficient.
Our simulations for the ZHC lattice
were more extensive, with at least $10^4$ realizations 
for systems, subject to both periodic
and twisted boundary conditions, of size $L=8,16,32,48$,
and at least $10^3$ realizations for systems of size $L=64,80$.
The largest systems here required $U=12$ to prevent the 
configuration from being affected by the boundaries of the lattice in
the $\hat u$ direction.

We have examined the displacement correlations of the 
minimal energy configurations
by computing the disorder averaged structure 
factor $\overline {S(\vec k)}$ of the displacement variables.  
This allows us to more clearly distinguish
the large length scale behavior; direct measurement of the width is more 
difficult to analyze due to finite size effects.  The orientationally
averaged structure factor $\overline {S(k)}$ 
has been obtained by averaging the value of
$\overline {S(\vec k)}$ over radial bins of 
size $\Delta k = 0.025$, and is presented in Fig.~\ref{fig:struct}.  
The error bars represent the fluctuations of $\overline {S(\vec k)}$ within 
each spherical shell.  
These fluctuations, which measure the 
anisotropy of the structure factor, generally decrease with decreasing $k$; 
for $k<0.75$, 
these fluctuations saturate at a value comparable to the statistical 
fluctuations in $S(\vec k)$
indicating the range in $k$ where the system is isotropic within the 
statistical fluctuations.
To extract the coefficient of the leading order divergence we have fit
$k^3\overline{S(k)}$ with the form
\begin{equation}
\label{eq:sk}
k^3\overline{S(k)} = A+Bk
\end{equation}
over the region $k<0.5$.  The leading order term of 
$\overline {S(k)} \sim k^{-3}$
indicates the quasi-long range order of the ground state.
For the ZHC lattice
we obtain $k^3 \overline{S(k)}=1.08(5)+1.02(6)k$; 
for the SHC lattice we obtain 
$k^3\overline {S(k)}=1.01(4)+0.46(4)k$.  
The error estimates on the parameters
in these fits represent the statistical uncertainty over the given fit 
range and systematic errors arising from the choice of the cutoff in $k$. 
Giamarchi and Le Doussal have applied a renormalization group technique
to order $\epsilon = 4-d$ to this system and obtained
$A=1.0$ \cite{gLeD}.  They have
also carried out a Gaussian variational calculation and obtained
$A=1.1$.  In both of these approximations the value of the coefficient 
is universal.

The form $Bk^{-2}$ of the leading order corrections to 
$\overline {S(k)}$ can be obtained
by a renormalization group calculation.  Upon renormalization, the 
periodic pinning potential
$V(u(\vec x),\vec x)$ introduces a new term into the Hamiltonian 
of the form $\vec \mu (\vec x)\dotprod \nabla u(\vec x)$ as when $d=2$ 
{\cite{CO,Fisher-2d}}.  
This random tilting field is short range correlated with 
$ \overline{ \vec \mu (\vec x) \dotprod \vec \mu(\vec y)} =
g^2\delta ^3(\vec x - \vec y)$.
Unlike the case where $d=2$, the strength of this field $g^2$ 
undergoes only a finite renormalization for $d=3$ \cite{Villain}.  
The effects of this term can be determined from the effective small 
length scale Hamiltonian
\begin{equation}
\label{eq:Hamiltionian-short}
H_{eff}=\int d^3x \ {c\over 2}[\nabla u(\vec x)]^2 + \vec 
\mu (\vec x) \dotprod \nabla u(\vec x)\ ,
\end{equation}
which ignores the periodic pinning potential.  Solving the 
equations of motion and averaging
over realizations of the tilting field predicts corrections of the 
form $Bk^{-2}$ with $B=g^2/c^2$
at $T=0$.  In order to consider the effects of the pinning potential 
and the tilting field
separately we relied on the fact that the renormalization group  
flow of $V(u(\vec x),\vec x)$ is unaffected by
the presence of a non-zero $g$ \cite{Villain,Fisher-2d}.  
The results of this analysis
can be confirmed by examining the stability of the functional
renormalization group
fixed point obtained by Giamarchi and Le Doussal in 
an $\epsilon = 4-d$ expansion \cite{gLeD}.
The most slowly decaying perturbation to the fixed point 
decays $\sim L^{-\epsilon}$, implying
corrections to the structure factor of the 
form $\sim k^{\epsilon -3}$. This order $\epsilon$
calculation also predicts the form of the corrections which are
observed in our simulation data.
     
Naturally, real-space measurements of the width must be consistent
with the structure factor.  We measured the disorder averaged 
squared width
\begin{equation}
\label{eq:width}
w^2= 
\overline {< \negthinspace u^2 \negthinspace >-
	<\negthinspace u \negthinspace >^2}\ ,
\end{equation}
where $<\ >$ denotes a spatial average over the system.
These data, shown in Fig.~\ref{fig:width}, have been fit 
using the real space version of Eq.\ (2): 
$w^2 = a + b \thinspace ln(L)+c/L\thinspace$.
The constant term arises from the short wavelength fluctuations,
while the second and third terms arise from the $k^{-3}$ and $k^{-2}$
terms in the structure factor respectively.  
The real space coefficient $b$ is related
to the leading order behavior of $S(k)$ by $b=A/4\pi^2 $.  
This three parameter fit gives
estimates for the values of $A$ and  $B$, the coefficients describing the 
long wavelength form of the
structure factor, which are consistent with that obtained 
by fitting $S(k)$ at a single system size.  
If the form of $S(k)$ depended on $L$, then this consistency would
not be maintained.  Direct comparison of the structure factor for 
various system sizes also demonstrates that $\overline {S(k)}$ exhibits 
negligible system size effects, 
except for the change in $k_{min}=L/2\pi$.

Other measures of the displacements provide us with additional information
on the structure of the ground state.  The 
disorder averaged extremal displacement difference,
$\Delta H = \overline {u_{max}-u_{min}}\thinspace$, 
Fig.~\ref{fig:extremes}, was found to grow logarithmically
with system size for both lattice types.  We computed least squares fits 
of the form $\Delta H = \tilde a + \tilde b \ ln(L)$ to obtain 
$\tilde b_{ZHC} =0.76(1)$ and $\tilde b_{SHC} = 0.70(1)$.
The coefficients
of the logarithmic term differ by less than $10$\% for the two lattices
studied here, suggesting that this measure of the system is weakly,
if at all, dependent on the lattice discretization of the medium.
This logarithmic growth is consistent with the following 
picture of the ground state
structure developed by Fisher \cite{lattice pinning}.  
At each length scale $R=b,\ b^2,\ b^3, \ldots$ the displacement
undergoes one shift of amount $\pm a$.  Furthermore the sign of the 
displacement shift is random at each scale, leading to the logarithmic 
growth of the
squared width.  When traversing from the minimum to the maximum the
signs of the displacements are strongly correlated, leading to a coherent 
sum, and a logarithmic dependence on the system size for the 
extremal differences results.

We have also determined the effect of coarse-graining the displacement
variable.  The coarse grained displacement is defined as the average
of $u$,
\begin{equation}
u_R(\vec y)= {1 \over R^3} \int_{\Omega_R(\vec y)} d^3x \ \ u(\vec x)\ ,
\end{equation}
over $\Omega_R(\vec y)$, a cube of size R centered at the point $\vec y$.  
We measured the fluctuations in these coarse grained
height variables $|\Delta u_R|^2= 
\overline {(u_R(\vec y) - u_R(\vec y+\vec b))^2}$,
with $\vec b$ the vector between the centers of the cubes which
touch at one corner.  This spatial averaging procedure 
is similar to a real space
renormalization transformation.  
Villain and Fernandez have explicitly carried
out a real space renormalization calculation for a 3-dimensional elastic
medium with cubic symmetry \cite{Villain}.  Their calculations indicate 
that $|\Delta u_R|^2$
has a finite limit as $R,L \rightarrow \infty$.  We have directly measured  
$|\Delta u_R|^2$ for the ZHC lattice (Fig.~\ref{fig:coarse}).
The coarse grained height fluctuations are
related to the structure factor by:
\begin{equation}
\label{eq:coarse}
|\Delta u_R|^2= \int_{BZ} {  {{d^3k}\over{(2\pi)^3}} \  
{|G(\vec k)|^2} \ {e^{ i \vec k \cdot \vec b}} \ {S(\vec k) }} 
\end{equation}
with $G(\vec k)=\int_{\Omega_R}{ d^3x \ \ e^{ i \vec k \cdot \vec x} }$. 
In the infinite volume limit $|\Delta u_R|^2$ depends only on the leading
order behavior, $Ak^{-3}$, of the structure factor and the limit of the
ratio $R/L$.  We 
have numerically evaluated the right hand side of Eq.~(\ref{eq:coarse}) 
in this limit
to obtain the infinite size limit presented in Fig.~\ref{fig:coarse}.  
For {\em finite sized systems}
the sub-leading order corrections to $S(k)$ contribute
to the coarse grained height fluctuations.  These terms lead to the 
divergence of 
$|\Delta u_R|^2$ as $R/L \rightarrow 0$.  The dominant corrections 
arise from the
$Bk^{-2}$ corrections to $S(k)$ seen in the structure factor data, 
but decay as $L^{-1}$.  The data are consistent with convergence to
a finite limit as $L \rightarrow \infty$.

For the ZHC lattice, we have also investigated the behavior of the system
subject to the twisted boundary conditions defined previously.  For each 
realization
of disorder we have compared the ground state energy with periodic boundary 
conditions, $E_p$, 
to that obtained with twisted boundary conditions along one of the
lattice directions, $E_t$.
This allows us to investigate the properties of the excitations induced by 
the change
in boundary conditions.  We identify the energy difference 
$E_{DW}=E_{t}-E_{p}$
with the energy of the domain wall.  The domain wall is identified
by the set of bonds that the interface intersects in one set of boundary 
conditions
but not the other.  Even though the domain wall could be identified by 
examining the
values of $u_{p}(\vec x)$ and $u_{t}(\vec x)$, it is more efficient to
identify the domain wall by examining the sets of cut bonds for 
each boundary conditions.
The cut bonds are projected along the $\hat u$ direction for each boundary 
condition.
The domain wall is then the symmetric difference between these two sets of 
bonds.
Without directly
simulating a dislocation loop itself, we are able to investigate 
the properties of the domain wall induced by 
the introduction of a single loop encircling the system.

The energetics of these domain walls dominate the random part of
the energy cost of introducing a dislocation loop into an elastic 
medium \cite{gingrashuse}.
The mean energy difference 
${\bar E_{DW}}$ grows linearly with the (linear) size $L$ of the domain 
wall (Fig.~\ref{fig:aveE}),
a result consistent with the scaling of the elastic contribution to the 
energy and the statistical symmetry of the disorder potential
of the continuum model\cite{Fisher}.  
We have also analyzed the variance $\sigma ^2(E_{DW})$
of the distribution of domain wall energies (Fig.~\ref{fig:varE}).  
No single power law fit of the form 
$\sigma^2 (E_{DW}) \propto L^{2 \theta}$ 
can adequately fit our data over the range of sizes simulated.
However the data are well described by the empirically determined form 
$\sigma ^2 (E_{DW}) = 0.031 L ^2 + 0.24 L$ displayed as the
solid line in Fig.~\ref{fig:varE}.  
The finite size correction
leads to a size dependent effective value of the 
exponent, $\theta_{eff}$, characterizing the scaling of the 
sample to sample fluctuations in the energy of low energy excitations.
Depending on the lower limit imposed on the fit,
single power-law fits $\sigma ^2 (E_{DW}) \sim L^{2 \theta}$
give $0.85 < \theta_{eff} < 0.92$ over this range of sizes.
The distribution
of domain wall energies depends on $\bar E_{DW}$ and $\sigma(E_{DW})$
only through the combination 
$\epsilon = (E_{DW} - \bar E_{DW} )/\sigma(E_{DW})$,
as can be seen in Fig.~\ref{fig:Edistribution}.
This collapsed distribution has more highly weighted tails
than a unit normal distribution.  The frequency with which negative energy
domain walls were observed in the simulations is significantly
higher that what one expects for a Gaussian distribution with the measured
mean and standard deviation at each system size 
(Fig.~\ref{fig:ProbabilityNeg}).
This behavior also occurs for high energy domain walls; to within 
the statistical uncertainty the distribution is symmetric about its
mean value.  Our data are consistent with both $\bar E_{DW}$, 
and $\sigma(E_{DW})$ increasing
linearly with $L$ for large systems.  
Fisher's argument assumed this behavior of the domain wall energetics 
in his domain wall
renormalization calculation indicating the marginal stability of the 
Bragg glass phase \cite{Fisher}.

Because of the balance between the elastic energy scale and the
scale of the energy fluctuations due to the disorder potential, 
the domain walls are highly convoluted
and expected to have a fractal dimension $d_f$ between 2 and 3.
Similar to the approach used for a 2-dimensional elastic medium 
{\cite{2delas,walls}},
we have measured the size of the domain wall by counting the number of 
bonds $N_b$ in the wall.  
The data for the area of the wall as a function of system size,
averaged over disorder,
can be fit by a simple power law, $N_b \sim L^{d_f}$,
with the fractal dimension of the domain wall  $d_f \approx 2.60$, 
shown in Fig.~\ref{fig:DW}(a).  This fit
has been taken after excluding the smallest system size $L=8$. 
However, the form of the residuals, see Fig.~\ref{fig:DW}(b),
indicate the presence of sub-leading
order corrections, suggesting that this may underestimate the 
value of $d_f$.  
In order to verify this, we have also fit the whole range of
data using the form $N_b = aL^{d_f}+bL^2$.  The second term arises from the
effectively two dimensional nature of the domain walls at 
small length scales.  This three parameter fit gives $d_f =2.65$; 
thus we conclude that our systematic
errors in estimating the fractal dimension are approximately $0.05$. 

In order to investigate other low energy excitations of this system, we 
have examined the sensitivity of the ground state to 
perturbations in the disorder potential.  Similar
studies have been carried out for disordered systems such as spin glasses
\cite{spin glass}, $(1+1)$-dimensional directed polymers in random 
media \cite{dprm}, and 2-dimensional elastic media \cite{2delas}.  
In all of these disorder dominated systems, perturbations
of relative strength $\delta$ in the disorder potential
introduce a length scale $L^* \sim \delta^{-1/\zeta}$,
$\zeta=d_f/2-\theta$, beyond which the ground state becomes 
uncorrelated with the reference ground  state.  
The exponent $\zeta$ characterizing the sensitivity of the ground state is
referred to as the chaos exponent.
These studies have been done by comparing the ground state for two 
correlated choices of the disorder potential.  
In our simulations, we have obtained 
the ground states for the two pinning potentials 
$V^{\pm}(u(\vec x),\vec x)=b(u(\vec x),\vec x) \pm d(u(\vec x),\vec x)$, 
with both terms periodic in the $\hat u$ direction.
The constant part of the potential $b(u(\vec x),\vec x)$ 
was an integer chosen uniformly from $[1000,2000]$.  
The term which generates differences between the realizations, 
$d(u(\vec x),\vec x)$, was chosen uniformly
from $[-d_{max}/2,d_{max}/2]$.  The parameter $\delta=d_{max}/2000$ 
characterizes
the relative strength of the perturbations.  This prescription was chosen to
ensure that for a fixed value of $\delta$ the distribution of the bond
weights is the same for both realizations of disorder.  
Our simulations include
values of $\delta$ ranging from 0.01 to 0.75, with at least $500$ 
independent realizations of
disorder at each $\delta$ and $L$.  
By performing scaling analysis of both the energetic and structural
correlations between the ground states for these realizations of disorder
we can extract the value of the chaos exponent.

We have found that both the energetic and structural correlations are
governed by the same length scale. 
First we calculated the domain wall energy $E^{\pm}_{DW}$ for the two 
disorder realizations $V^{\pm}$,
which can then be used to compute the domain wall energy correlation
function\cite{spin glass}
\begin{equation}
G={(\overline{E_{DW}^{+}-\overline{E_{DW}^{+}}})\ \ 
\overline{(E_{DW}^{-}-\overline{E_{DW}^{-}})}\over{ \sigma ( E_{DW}^{+} )
\sigma (E_{DW}^{-}) }\ }.
\end{equation}  
The simple scaling form $G=f(\delta L^{\zeta})$ describes  our data 
well (Fig.~\ref{fig:correlation}).
We found reasonable data collapse for $\zeta = 0.38(4)$, taking into account
the statistical errors.  
The value of the chaos exponent can be related to the domain wall 
fractal dimension
and the energy fluctuation exponent by a simple scaling argument as in 
the case
of spin glasses \cite{spin glass}.  The perturbations introduce a 
random change in the energy of the domain wall of order 
$\delta L ^{d_f/2}$ because the
perturbations are uncorrelated with the location of the domain wall.
The typical fluctuations in the domain wall energy scale as 
$L^{ \theta}$.
When these energy scales become comparable,  at a length scale $L^* \sim 
\delta^{-1/\zeta}$,
the domain wall energies become uncorrelated.  
When using the effective value of the energy fluctuation exponent,
$\theta_{eff} \approx 0.9$, this scaling relation holds to within
5\% accuracy.

We can understand the structural deformations induced by the bond 
perturbations by reasoning similar to that for domain wall correlations.  
Here we consider the differences in the ground states of the system
(with periodic boundary conditions) due to the changes in the 
pinning potential.  Again, the ratio of the
energy change due to the random perturbations, and that of the fluctuations
in the energy landscape, $\delta L^{d_f/2} / L^{\theta}$, determines the
behavior of the system.  In response to Zhang's simulation of directed
polymers in random media, Feigel'man and Vinokur had argued 
that the probability of a positional excitation
grows linearly with $\delta L^\zeta$ for small values of the perturbation
strength \cite{dprm}.  
Following their argument, we expect the probability of a 
change in the displacement variable at a single location
to grow linearly with $\delta L^\zeta$.  Unlike the case for the 
directed polymer,
the magnitude of the differences is bounded for periodic pinning since 
excitations with $|\Delta u| > a$
probe the same energy landscape as those with $|\Delta u| < a$.  
Thus, the spatially
and disorder
averaged mean squared displacement difference
$\chi=\overline{ 
< \negthinspace (u^{+}(\vec x)-u^{-}(\vec x))^2 \negthinspace >
}  \sim \delta L^\zeta$ for small perturbations.
For large values of $\delta$ the ground states are completely 
decorrelated, and we expect that $\chi \sim ln(L)$.  We have found that
the data for $\chi$ collapse according to the scaling form
$\chi = f( \delta L^\zeta)$, with $\zeta=0.39(2)$
(Fig.~\ref{fig:spatial}).
Our data are consistent with the results obtained by the scaling 
argument in both limiting cases. 
Before computing $\chi$,
we made the transformation $u^{+} \rightarrow u^{+} +na$, where $n$
is the integer which maximizes the number of locations at which 
$u^{+}(\vec x) - u^{-}(\vec x) = 0$ for each realization.  
This transformation minimizes $\chi$ over the discrete set of 
global translations which leave the energy of the medium invariant.  
Our scaling ansatz is significantly different from that proposed for
the 2-dimensional elastic medium $\chi = \delta^{1/\zeta}f(L)$ \cite{2delas}.  
This form cannot adequately collapse our data over the range of 
parameters we have simulated.

Implicit in this discussion
is the assumption that both the perturbation induced deformations, and the
boundary condition induced domain walls are characterized by the same 
fractal dimension.  
Even for relatively small perturbations in the disorder, the 
deformations are typically composed of a set of disconnected clusters.  
We have directly measured these clusters' fractal dimension.
The size of a cluster $R$
is defined as the average of the sides of the bounding box which encloses
the cluster and is measured in units of the lattice spacing.  Our algorithm
identifies the sets of nodes on which $u^{+}(\vec x)-u^{-}(\vec x) \neq 0$ 
after performing the translation which minimizes $\chi$.
The surface area of a cluster $s$ is the number of nodes with
neighbors not in the cluster.
We have collapsed the data for $\delta = 0.05$ using the finite 
size scaling form $s=L^{d_f}f(R/L)$ (Fig.~\ref{fig:clusterDim}). 
We expect that the scaling function $f(R/L)$ should
have the form $f \sim (R/L)^{d_f}$ in the region $R/L << 1$, $R >> 1$,
but this regime is not clearly visible in Fig.~\ref{fig:clusterDim} due to 
lattice and finite size effects.
Despite this, the best collapse of the data, for $0.5 < R/L < 1.0$,
provides an estimate $d_f = 2.65(10)$ which is consistent with the estimate
for the domain wall fractal dimension.  Similar analysis at other values 
of $\delta$ provide equivalent values for the cluster fractal dimension.  
The anomalous data at $R/L \approx 1$
arise from the rare clusters which span the system in all directions.  
This scaling also
breaks down for clusters with  $R < 8$, where lattice effects make the
surface effectively 2-dimensional (Fig.~\ref{fig:smallCluster}).    
The equality of the fractal dimension of the boundary condition induced 
domain walls
and bond perturbation induced deformations can be justified by
a simple argument.  For small values of the disorder perturbation
parameter, the cluster boundaries lie in regions where there is a
small energy cost to deforming the medium.  If one considers only
a small volume containing a portion of the cluster boundary,
the structural difference is the same as would be caused
by the change from periodic to twisted boundary conditions on that
volume.  Thus both the deformations induced by small changes
in the disorder potential and those caused by a change in boundary
conditions should be characterized by the same fractal dimension.
Despite the fact that
the scaling regime is inaccessible due to the limits on the size
of systems studied, our data are consistent with the conclusion
that the fractal dimension of the clusters which compose the
deformations is the same as that of the boundary condition induced
domain walls.

We have performed extensive numerical simulation of a model 3-dimensional 
elastic medium subject to quenched disorder with scalar discrete 
displacements.  Our results for the structure in the Bragg glass phase
indicate that the structure factor has divergences of the 
form $S(k) \sim Ak^{-3}$. 
Our results for the coefficient A fall between the 
approximate values $A=1.0$, and $A=1.1$,
obtained via a renormalization group and a replica approach \cite{gLeD}.  
The observed energetics of the boundary condition induced domain walls
indirectly support arguments for the stability of the
Bragg glass phase.  These domain walls correspond to the 
elastic deformations due to the introduction of a single dislocation 
loop winding
around the system.  Our data are consistent with the hypothesis that the
mean energy and the energy fluctuations
of a section of domain wall both scale linearly with the 
linear size of the section for large sizes.  
This balance is a crucial element of the analysis
carried out by Fisher indicating the marginal stability of
the Bragg glass phase to the introduction of dislocations \cite{Fisher}.
We are also able to measure the spatial structure
of these domain walls and obtain their fractal dimension $d_f=2.60(5)$.
We have observed that random changes in the disorder potential of
relative strength $\delta$ decorrelate the ground state on length scales
larger than $L^* \sim \delta ^{-1/\zeta}$ with $\zeta=0.385(40)$.
The properties of the domain walls and this sensitivity to disorder
perturbations can be related to each other by the scaling relation
$\zeta = d_f/2 - \theta$,
where $\theta$ characterizes the fluctuations in the low energy
excitations.




\input epsf.tex

\def\pmb#1{\setbox0=\hbox{#1}%
  \kern-.025em\copy0\kern-\wd0
  \kern.05em\copy0\kern-\wd0
  \kern-.025em\raise.0433em\box0 }
\def\grad{{\pmb{$\nabla$}}}

\draft

\begin{figure}

\caption{
 Numerically calculated structure factor for a pinned elastic medium, 
 averaged over disorder realizations.  Data is shown for:
 (a) the Z-centered  hypercubic lattice for  L=80 and  1000 realizations,
 and (b) the simple hypercubic lattice, L=80, 1000 realizations.
 The data have been coarse grained into bins of size $\Delta k=0.025$,
 and only every third data point is displayed. 
 The error bars
 represent the fluctuations in the values of $\overline {S(k)}$ 
 within each bin, and
 hence are a measure of the anisotropy of $\overline {S(k)}$.
 For both lattices a fit 
 of the form $k^3\overline {S(k)}=A+Bk$ has
 been taken over the region $k \le 0.5$.  
 The resulting least squares fits are shown as solid lines.
 For wave vectors in this region
 the anisotropy of the structure factor is comparable to the size of the
 statistical fluctuations in the value of $S(k)$.
 The coefficient of the leading order divergent term, $Ak^{-3}$, 
 can be extracted from these fits:  $A_{ZHC}=1.08(5)$, \break 
 $A_{SHC}=1.01(4)$, where the errors include both the statistical 
 errors and our estimate of the systematic errors.
}
\label{fig:struct}

\end{figure}

\begin{figure}
\caption{
	Disorder averaged roughness of the interface representation 
	of the elastic medium.
	The statistical errors at each size are comparable to, or 
	smaller than the plot points.
	Fits of the form $a+b\ ln(L)+c/L$, which include the form of the
	finite size corrections indicated by the structure factor data, 
	are represented by the solid lines.
	The coefficient $b$ in these fits is related 
	to the coefficient of the divergence
	of the structure factor by $b=A/4\pi^2$.  These fits provide 
	values of $A_{SHC}=1.03(6)$, $A_{ZHC}=1.02(1)$.
	}
\label{fig:width}
\end{figure}

\begin{figure}
\caption{
	Disorder averaged extremal displacement difference, 
	$\Delta H=\overline {u_{max}-u_{min}}\thinspace$, 
	as a function of system size.
	The statistical errors at each size are smaller than the
	plot points.
	Over this range of sizes the behavior is logarithmic,
	as indicated by the solid lines which are fits of the form
	$\Delta H = \tilde a + \tilde b \ ln(L)$.
	The least squares values for the coefficient of 
	the logarithmic term are
	$\tilde b_{ZHC}=0.76(1)$, $\tilde b_{SHC}=0.70(1)$.
}
\label{fig:extremes}
\end{figure}

\begin{figure}
\caption{
	Behavior of the coarse grained displacement differences.  
	$|\Delta u|$ is the difference between the average 
	displacement of the medium
	between two adjacent cubic regions of size R, averaged over 
	disorder.  
	The statistical errors at each size are smaller than the
	plot points; the dashed lines serve as guides to the eye.
	The solid line represents the
	$L\rightarrow \infty $ extrapolation for the coarse grained 
	displacement 
	differences and was calculated using the long wavelength
	behavior of the structure factor.
	}
\label{fig:coarse}
\end{figure}

\begin{figure}
\caption{
   	Average energy difference between the ground state of the system 
	with periodic and twisted 
	boundary conditions.  The data are well fit by
	$E_{DW} = 0.395(1)L$.
}
\label{fig:aveE}
\end{figure}

\begin{figure}
\caption{
   	Variance of the energy difference $E_{DW}$ 
	between the ground states  with periodic and twisted
	boundary conditions.  The fit, $\sigma ^2 = 0.031(1)L^2+0.24(1)L$,
	includes an empirically estimated  correction to scaling.
}
\label{fig:varE}
\end{figure}

\begin{figure}
\caption{
	Data collapse for the distribution of 
	$\epsilon =(E_{DW}-\bar E_{DW})/\sigma(E_{DW})$, 
	the normalized energy difference between the 
	ground states of the system with twisted and
	periodic boundary conditions.  The points represent the observed 
	frequencies of $\epsilon$ over bins of size 0.2 for at least 
	$10^4$
	realizations at each system size.  
}
\label{fig:Edistribution}
\end{figure}

\begin{figure}
\caption{
	Frequency with which negative energy domain walls 
	were observed.  The solid line
	represents the expected probability given the assumption
	that the distribution is a Gaussian with the observed 
	mean and standard deviation.
}
\label{fig:ProbabilityNeg}
\end{figure}

\begin{figure}
\caption{
	Structural difference between systems with periodic and twisted
	boundary conditions.  Part (a) displays the area $N_b$ of the
	domain wall as a function of the system size $L$.  The statistical
	errors at each size are smaller than the plot points.  The solid
	line represents a fit over the region $L \geq 10$ with the
	form $f(L)=aL^{d_f}$.  The least squares fit provides an 
	estimate of the domain wall fractal dimension $d_f=2.60$.  However
	the form of the residuals for this fit, 
	part (b), suggest the presence of small sub-dominant corrections.
}
\label{fig:DW}
\end{figure}

\begin{figure}
\caption{
	Boundary condition induced domain wall energy correlation
	function $G$.  The scaling variable is a combination of 
	the system size $L$ and bond perturbation strength $\delta$.
	The typical magnitude of the of these perturbations
	ranges from $\delta = 0.01$ to $\delta = 0.75$ when measured
	in units of the range of the (unperturbed) pinning potential.
	Each data point represents an average over at least $500$
	realizations of disorder.  
	The data collapse  provides 
	an estimate for the chaos exponent $\zeta=0.38(4)$;
	the error was estimated by determining the range in
	$\zeta$ where the collapse is reasonable.
}	
\label{fig:correlation}
\end{figure}

\begin{figure}
\caption{
	Configurational differences induced by perturbing the 
	disorder potential randomly at
	each location with relative strength $\delta$.  
	The response function, $\chi=
	\overline{V^{-1}\sum_{\vec x} (u^{+}(\vec x)-u^{-}(\vec x))^2}$, 
	is the spatially averaged squared displacement 
	difference between the reference and perturbed ground states.
	The data have the scaling form $\chi=f(\delta L^\zeta)$, 
	with $\zeta=0.39(2)$.
}
\label{fig:spatial}
\end{figure}

\begin{figure}
\caption{
	Scaling of the surface area of the singly connected components
	of deformations induced by perturbations of relative strength
	$\delta = 0.05$ in the disorder potential.  The data
	collapse for $d_f = 2.65$, with an uncertainty of $\pm 0.1$.
	As a guide to the eye, the $d_f=2.65$ power law behavior
	of the average surface area, $s$,  as a function of the cluster's
	radius, R, is indicated by the solid line.  Because each
	realization of disorder generates numerous clusters of
	various sizes, the statistical errors for the average
	surface area are smaller than the plot points.
	}
\label{fig:clusterDim}
\end{figure}

\begin{figure}
\caption{
	Relationship between the surface area and size of
	small connected clusters of deformation induced by small 
	changes in the
	disorder potential.  For $R < 8$ their surfaces
	are approximately two dimensional and independent 
	of the system size, 
	as indicated by the solid line representing $s=R^2$.
	This behavior does not persist to larger clusters.  In order
	to illustrate the cross over, the large cluster behavior
	$s \sim R^{2.65}$ is represented by the dotted line.
	The statistical errors for each data point are 
	smaller than the plot points.
	}
\label{fig:smallCluster}
\end{figure}


\begin{references}
\bibitem{ZC}
This choice ignores the 12 nearest neighbor bonds which have a zero 
projection along $\hat u$.
\bibitem{lattice types}
The displacement is defined on a diamond structure lattice 
for the SHC lattice and a simple cubic lattice for the ZHC lattice.
\end{references}
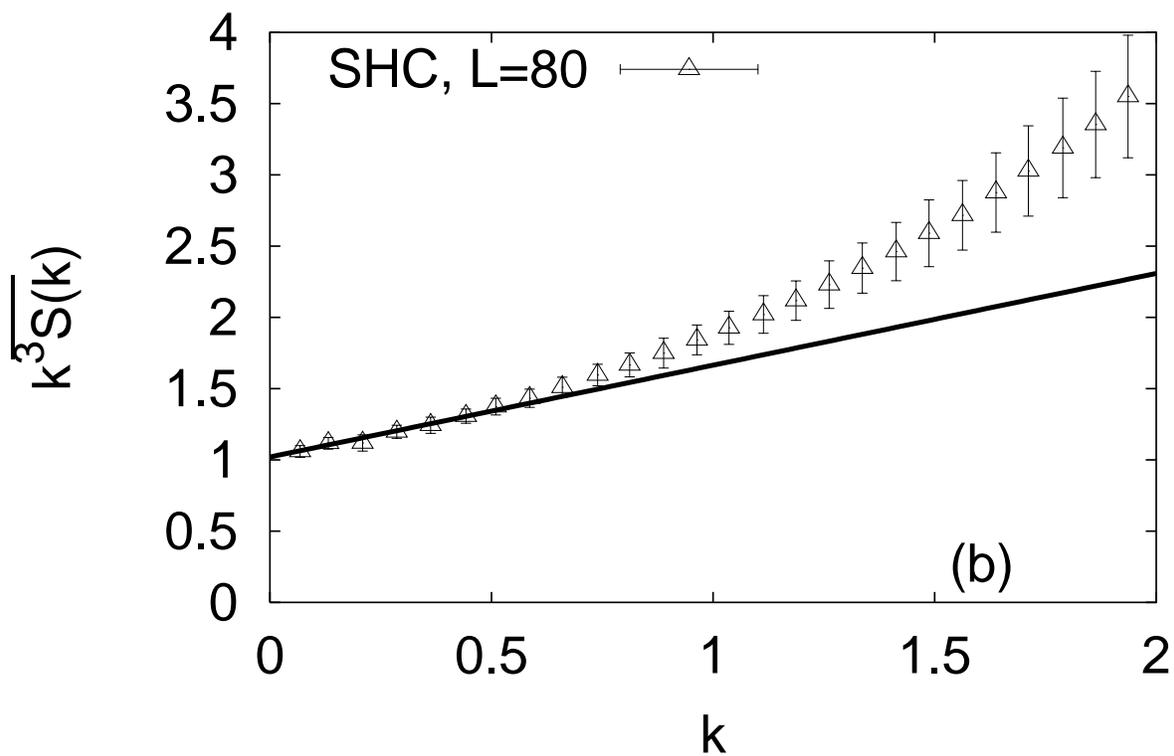
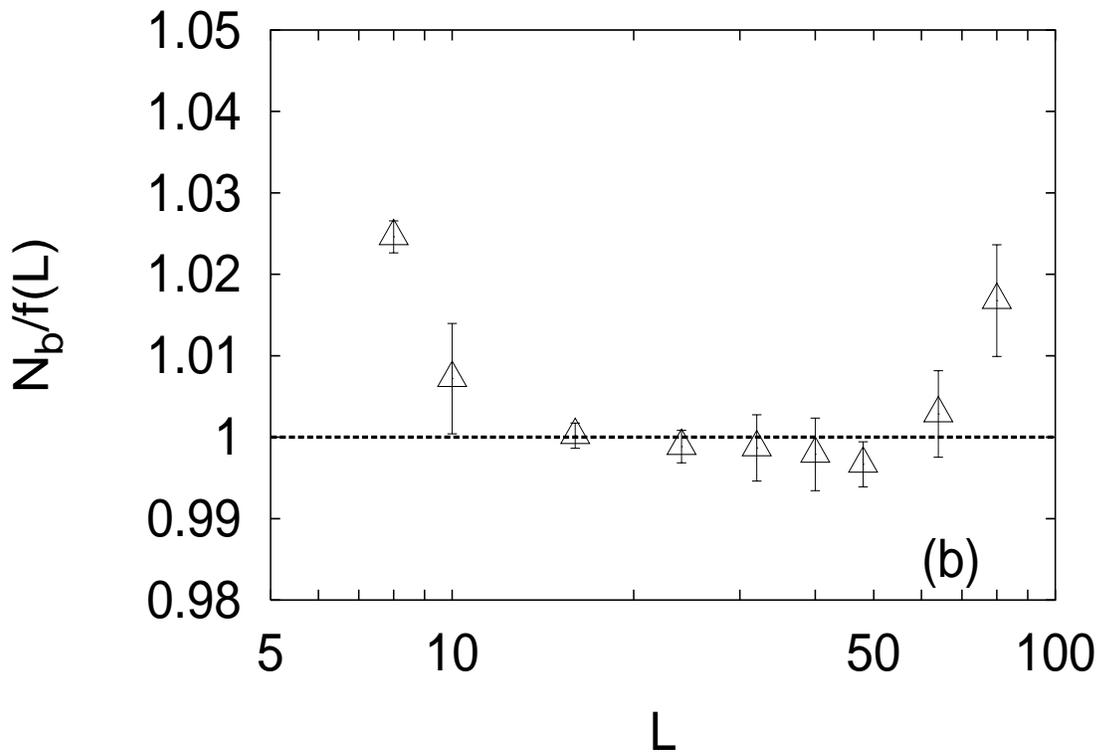
\end{document}